\documentclass[twocolumn, showpacs]{revtex4}

\usepackage{graphicx}
\usepackage{dcolumn}
\usepackage{bm}

\begin{document}


\title{Evolution of
particle-scale dynamics in an aging clay suspension}

\author{R.
Bandyopadhyay$^{1}$, D. Liang$^{1}$, H. Yardimci$^{1}$, D. A. Sessoms$^{2}$, M.
A. Borthwick$^{3}$, S. G. J. Mochrie$^{4}$, J. L. Harden$^{2}$ and
R.
L. Leheny$^{1}$}

\affiliation{$^{1}$Department of Physics and
Astronomy, Johns Hopkins
University, Baltimore, MD 21218, USA.
$^{2}$Department
of Chemical and Biomolecular Engineering, Johns Hopkins
University,
Baltimore, MD 21218, USA.  $^{3}$Department of Physics,
Massachusetts
Institute of Technology,
Cambridge, MA 02139, USA.  $^{4}$Department of
Physics, Yale University, New Haven, CT 06520,
USA}

\date{\today}

\begin{abstract}
\noindent Multispeckle x-ray photon correlation spectroscopy
was employed to characterize the slow dynamics of a colloidal
suspension formed by highly-charged, nanometer-sized disks.  At scattering wave
vectors $q$ corresponding to interparticle length scales, the dynamic
structure factor follows a form $f(q,t) \sim \exp[-(t/\tau)^{\beta}$],
where $\beta \approx$ 1.5.  The characteristic relaxation time $\tau$
increases with the sample age $t_a$ approximately as $\tau \sim
t_a^{1.8}$ and decreases with $q$ approximately as $\tau \sim q^{-1}$.  Such a
compressed exponential decay with relaxation time that varies inversely
with $q$ is consistent with recent models that describe the dynamics in
disordered elastic media in terms of strain from random, local
structural rearrangements.  The amplitude of the measured decay in $f(q,t)$
varies with $q$ in a manner that implies caged particle motion at short
times.  The decrease in the range of this motion and an increase in
suspension conductivity with increasing $t_a$ indicate a growth in the
interparticle repulsion as the mechanism for internal stress development
implied by the models.
\end{abstract}


\maketitle

A signature feature of many disordered materials is the protracted
evolution of their dynamic and thermodynamic properties.  This
process, known as aging, appears in a variety of systems including
polymers~\cite{polymer}, spin glasses~\cite{spinglass}, molecular
glasses \cite{leheny_epl},
and colloidal gels~\cite{cipelletti}.  The observation of common
behavior among seemingly disparate materials indicates generic underlying
mechanisms for aging that have attracted strong theoretical interest into
the nature of this out-of-equilibrium phenomenon
\cite{cugliandolo,bouchaudbook,franz,sollich}. Experiments that probe aging have typically
measured the temporal evolution of response functions, such as magnetic
susceptibilities or elastic moduli, and far fewer experiments have
characterized dynamical correlation functions in aging
systems~\cite{cipelletti,cipelletti_faraday,bellour,abou}.  However, because measurements of
correlation functions generally access wave-vector dependence, they
provide insight into variations in dynamics with length scale that can
illuminate the microscopic origins of aging.  To address this issue, we
have conducted x-ray photon correlation spectroscopy (XPCS) studies on
aqueous clay suspensions to characterize their intermediate scattering
function during aging.   The large wave vectors accessible with XPCS make
these studies among the first to investigate the effects of aging on
dynamics at interparticle length scales~\cite{weeks}.

The system under
study is laponite XLG (Southern Clay Products), a synthetic Hectorite
clay composed of discoidal particles
1 nm wide and 15 nm in radius. When
dispersed in deionized water, the particles have a net
negative charge
leading to a repulsive colloidal suspension~\cite{phasediagram,
repulsive}.
At volume fractions $\phi \geq\phi^*\simeq 0.007$, laponite
suspensions gradually transform into a soft solid with thixotropic
response
to stress~\cite{thixo} and anomalous rheology~\cite{phasediagram}.
In
addition, light scattering experiments have demonstrated that the long
length-scale
dynamics in such suspensions displays characteristic aging
behavior \cite{bellour,abou}.
To study the evolution of particle-scale
dynamics using XPCS, we prepared a suspension with $\phi$ = 0.012 by
mixing oven-dried clay in deionized water.  The resulting solution had pH
= 9.8 and ionic strength $\simeq 2\times10^{-4}$ M.  The clarified
solution was filtered through 0.45 $\mu$m pores to break up undissolved
aggregates, and the age of the sample $t_a$ was measured from this
filtering time.

Measurements were performed at sector 8-ID-I of the Advanced Photon
Source (APS).  Details regarding the beam line optics employed to create a
partially coherent x-ray beam have been presented
elsewhere~\cite{lumma_pre}.  The laponite suspension was contained in a sealed holder 700
$\mu$m thick with thin kapton windows for transmission scattering.  The
scattering intensity was recorded by a direct-illuminated CCD area
detector (Princeton Instruments EEG model 37) 3.4 m after the sample
covering a range of wave vectors $q$ from 0.05 nm$^{-1}$ to 0.3 nm$^{-1}$.  At
different sample ages a series of scattering images were recorded to
determine the ensemble-averaged intensity autocorrelation  function
$g_2(q,t) = \frac{\langle I_{ij}(q,t_a)I_{ij}(q,t_{a}+t)\rangle_{ij}}{
\langle I_{ij}(q,t_a)\rangle_{ij}\langle
I_{ij}(q,t_{a}+t)\rangle_{ij}}$,
where $\langle...\rangle_{ij}$ denotes an average over
pixels~\cite{lumma_pre}. The minimum delay time $t$, set by the transfer rate of the CCD,
was 1.6 s.  The longest time was set at 1000 s to avoid effects due to
limitations in measurement stability that led to artificial decays in
$g_2(q, t)$ at several thousand seconds.  Measurements of $g_2(q,t)$
were made at ages from $t_a = 1.3\times10^{4}$ s to $t_a = 2\times10^{5}$
s, during which time the sample evolved from a viscous liquid to a
thixotropic solid capable of supporting its own weight.

Figure 1 shows results for $g_2(q, t)$ at $q = 0.14$ nm$^{-1}$
measured at several $t_a$. The autocorrelation function displays a
relaxation with an amplitude and characteristic decay time that
increase with increasing age.  The dynamic structure factor
$f(q,t)$ can be calculated from $g_2(q, t)$ using the Siegert
relation. Modeling $f(q,t)$ with a stretched exponential lineshape
$f(q,t) = Aexp{(-t/\tau)}^{\beta}$, the intensity autocorrelation
function has the form
\begin{equation}
 g_2(q, t) =  1+b[Aexp[-(t/\tau)^{\beta}]]^2
\end{equation}
\noindent where $A$ is the short-time ($t <$ 1 s) plateau amplitude of
$f(q,t)$, $b$ is the Siegert factor, $\tau$ is a characteristic
relaxation time, and $\beta$ is the stretching exponent that characterizes the
lineshape.  The solid lines in Fig.~1 are the results of fits to
Eq.~(1), which describes the data accurately over the full range of $q$ and
$t_a$.

\begin{figure}
\includegraphics[scale=1.0]{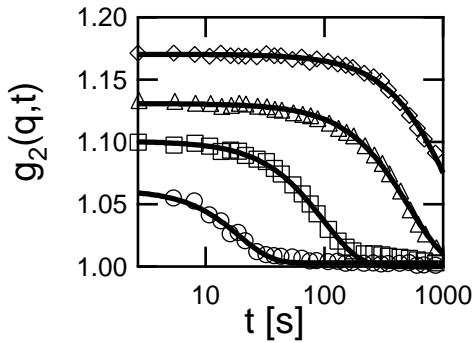}
\caption{Intensity autocorrelation function $g_{2}(q,t)$ at
$q=0.14$ nm$^{-1}$ for a laponite suspension of volume fraction
$\phi = 0.012$ at four ages:  $t_a$ = 1.3$\times$10$^{4}$ s
(circles), 3$\times$10$^{4}$ s (squares), 9$\times$10$^{4}$ s
(triangles) and 2$\times$10$^{5}$ s (diamonds).  Solid lines are
the results of fits to Eq.~(1). }
\end{figure}

Results for the relaxation time $\tau$ and the amplitude of the
decay in $g_{2}(q,t)$, given by $bA^2$, at $q=0.14$ nm$^{-1}$ are
shown as a function of sample age in Figs.~2(a) and 2(b),
respectively.  Over the measured range in $t_a$, $\tau$ scales
with age approximately as $\tau \sim {t_a}^{\mu}$ with $\mu = 1.8
\pm 0.2$.  While such power-law scaling between $\tau$ and $t_a$
is a generic feature of aging systems, typically $\mu \leq 1$, and
an exponent $\mu$ greater than one is unusual. Indeed, aging with
$\mu$ = 1 has been observed with light scattering for a laponite
suspension at $\phi$ = 0.0136 for $t_a >$ 10$^{4}$
s~\cite{bellour}. However, light scattering measurements observe a
much stronger relation at early ages, specifically $\tau \sim
exp(t_a)$~\cite{bellour,abou}.  Conceivably, this rapid evolution
of dynamics at early ages could extend to larger $t_a$ for smaller
length scales, leading to an apparent scaling exponent of $\mu >
1$.  For the range of ages at which we observe $\mu > 1$, however,
$g_{2}(q,t)$ possesses a faster-than-exponential decay
characteristic of solid-like behavior and thus does not support
this picture.  Instead, we interpret the observed scaling with
$\mu >$ 1 as a potentially general feature of aging dynamics at
interparticle length scales.

Specifically, the shape of $g_{2}(q,t)$ is inconsistent with a simple
exponential relaxation $\beta = 1$, except perhaps at the earliest age
$t_a= 1.3\times10^4$ s, and best fits to Eq.~(1) give $\beta \approx
1.5$, with no systematic variation with age or $q$.  Such a compressed
exponential decay is incompatible with diffusive, fluid-like motion of the
particles.  While such hyperdiffusive relaxation with $\beta >1$ may
seem unusual, similar behavior with $\beta = 1.5$ has been observed at
smaller wave vectors with dynamic light scattering on a number of soft
solids including colloidal gels~\cite{cipelletti}, clay
suspensions~\cite{bellour}, micellar polycrystals~\cite{cipelletti_faraday}, and
concentrated emulsions~\cite{cipelletti_faraday}.   Recently, Cipelletti {\it
et al.}~\cite{cipelletti_faraday}, have advanced a microscopic picture
for this dynamics, describing it in terms of the ballistic motion of
elastic deformation in response to heterogeneous local stress.  A
specific model by Bouchaud and Pitard~\cite{bouchaud} associates local
rearrangments, or ``micro-collapses'', of particles with the source for
dipolar stress fields.  A key feature of such dynamics is an inverse relation
between wave vector and relaxation time, $\tau \sim q^{-1}$.    As
shown in Fig.~3, $\tau$ for the laponite suspension varies inversely with
$q$, $\tau \sim q^{-0.9 \pm 0.1}$, over the range of wave vectors
covered in the XPCS measurement in agreement with this prediction.  This
scaling holds for all $t_a$ where $\beta \approx 1.5$.

\begin{figure}
\includegraphics[scale=1.0]{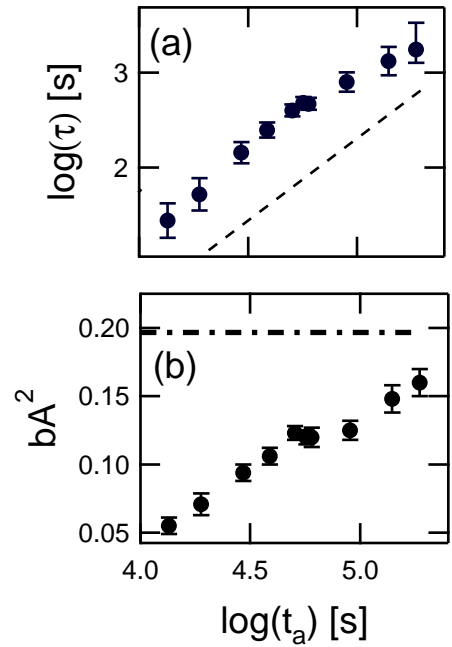}
 \caption{(a) Characteristic relaxation
time and (b) amplitude of $g_{2}(q,t)$ at short times for a
laponite suspension of $\phi = 0.012$ at $q=0.14$ nm$^{-1}$ as a
function of age.  The dashed line in (a) has a slope of 1.8.   The
dashed-dotted line in (b) shows the value of the contrast measured
using a static aerogel sample.}
\end{figure}

The observations with XPCS of $\beta \approx 1.5$ and $\tau \sim
q^{-1}$ for laponite thus extends quite unexpectedly this solid-like dynamics
into the particle-scale regime.   In particular, combined with previous
small angle dynamic light scattering measurements on colloidal gels at
very small wave vectors~\cite{cipelletti}, these results demonstrate
that such hyperdiffusive relaxation can occur at length scales varying
nearly five orders of magnitude relative to particle size.   The
relaxation mechanism suggested by Cipelletti {\it et al.}, in which particles
and their neighbors translate together with some velocity, provides a
natural explanation for this broad range.  In the absence of other
dynamics, such motion leads to $\tau \sim q^{-1}$ for all $q$ greater than
the inverse of the distance that particles travel under the strain.

We emphasize that the slow local dynamics of the laponite suspension
contrasts with structural dynamics in glass-forming liquids.  In glassy
liquids, the main relaxation, known as the ``alpha'' relaxation,
displays de Gennes narrowing at interparticle length scales such that $\tau
q^2$ varies proportionally with the structure factor~\cite{segre,tolle}.
We have investigated in detail the local structure in laponite
suspensions through small angle neutron scattering (SANS) at the NIST Center
for Neutron Research.  The scattering intensity $I(q)$, also shown in
Fig.~3 for a suspension of $\phi = 0.012$, displays a weak interparticle
structure factor peak near $q = 0.13$ nm$^{-1}$ (corresponding to a
center-to-center particle distance of $2\pi/0.13 \approx 48$ nm for these
30 nm diameter disks).  At higher $q$, $I(q)$ decays as $I(q) \sim
q^{-2}$, characteristic of the disk form factor.  Unlike with de Gennes
narrowing, $\tau$ does not track the peak in $I(q)$.  The insensitivity of
$\tau$ to the interparticle correlations demonstrates conclusively that
the slow dynamics of laponite are distinct from the collective
diffusion of glassy systems.

\begin{figure}
\includegraphics[scale=1.0]{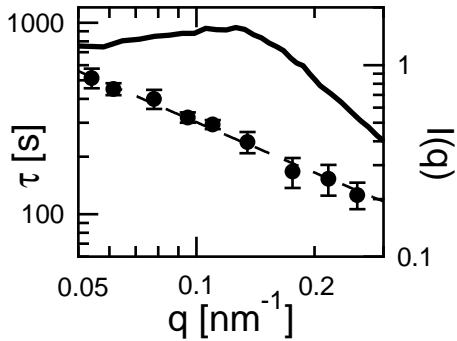}
 \caption{Characteristic relaxation time
$\tau$ (circles) at $t_a= 4\times10^{4}$ s and scattering
intensity $I(q)$ (solid line) from SANS for a laponite suspension
with $\phi=0.012$  as a function of wave vector $q$.  The dashed
line displays the fit result $\tau \sim q^{-0.9 \pm 0.1}$.}
\end{figure}

While the picture that Cipelletti {\it et al.} introduce accounts well
for these slow dynamics, an obvious question is the source of local
stress in these soft solids.  This issue is especially germane to
laponite, since the suspensions solidify over the course of hours or days.
Thus, unlike for systems quenched into a state far from equilibrium,
laponite suspensions likely enter the solid phase annealed and free of
residual stress.  In particular, the micro-collapse of
particles~\cite{bouchaud} is unlikely in laponite for which the interparticle interactions
are primarily repulsive at high pH and low ionic
strength~\cite{phasediagram,repulsive,morvan}.  However, through the fast local dynamics
implied by the XPCS results, we identify a microscopic mechanism,
specifically the growth of the interparticle repulsion with $t_a$, by which
stress can develop.

Information regarding the fast dynamics derives from the amplitude of
$g_{2}(q,t)$ at short times.  As shown in Fig.~2(b), this amplitude
increases steadily with $t_a$.  Ideally, $g_{2}(q,t) \rightarrow 2$ as $t
\rightarrow 0$.  Values less than two can reflect dynamic processes at
times shorter than those experimentally accessible causing an apparent
suppression in the structure factor amplitude, $A <$ 1.  In addition,
instrumental effects can lead to a Siegert factor $b <$ 1.  To assess the
instrumental effects, we compare the short time amplitude of
$g_{2}(q,t)$ for laponite with that measured for a static aerogel sample. The
amplitude of $g_{2}(q,t)$ for aerogel, shown by the dashed-dotted line in
Fig.~2(b), gives the Siegert factor $b \simeq$ 0.20, independent of $q$
and consistent with a value expected based on the x-ray optics. The
suppression in the amplitude of $g_{2}(q,t)$ for the laponite suspension
with respect to that of aerogel thus implies a partial decay to a
plateau at inaccessibly short times and consequently to a two-step
relaxation.  Such two-step relaxations occur commonly in disordered systems
including dense colloidal suspensions, polymer and colloidal gels, polymer
solutions and supercooled liquids.  In the case of supercooled liquids,
the first part of this decay, termed the ``beta'' relaxation,
represents the caged motion of a molecule in the confined space defined by its
neighbors. Within this picture of caged motion, the short time plateau
value is given by the Debye-Waller factor,
\begin{equation}
A =
\exp(-{q}^{2}\langle\bar{r}^{2}\rangle/3).
\end{equation}
\noindent Fig.~4(a) displays ln$(A)$ versus $q^2$
for the laponite suspension at various $t_a$.  The observed linear
relationships support the picture of caged motion of the laponite particles
at short times.  Fig.~4(b), which shows values of the root mean squared
displacements $\langle\bar{r}^{2}\rangle^{1/2}$ extracted from linear
fits, reveals that this rapid motion becomes increasingly restricted
spatially as the system ages.

We identify this increased restriction with
an evolution in the interparticle potential.  As mentioned above, the
laponite become charged in aqueous solution, specifically via the
unbinding of clay particles and subsequent dissociation of Na$^{+}$ ions from
exposed faces, resulting in a screened Coulombic repulsion.  To obtain
evidence about this interaction, we determined the d.~c.~conductivity
$\sigma$ of a suspension with $\phi = 0.012$ as a function of $t_a$ by
measuring the low-frequency impedance of a capacitor filled with the
suspension.  As shown in Fig.~4(b), $\sigma$ increases with increasing
$t_a$.  While the observed increase is modest, it is fully reproducible
and far exceeds the uncertainty in the impedance measurements, which was
roughly 0.1\%~\cite{sigma_error}.    Further, the relation between
$\sigma$ and ionic concentration is highly nonlinear for low conductivity
suspensions~\cite{sen}.  Hence, the small fractional increase in
$\sigma$ implies a much larger change in ionic double-layer and, therefore, in
repulsive interaction between particles.

\begin{figure}
\includegraphics[scale=0.9]{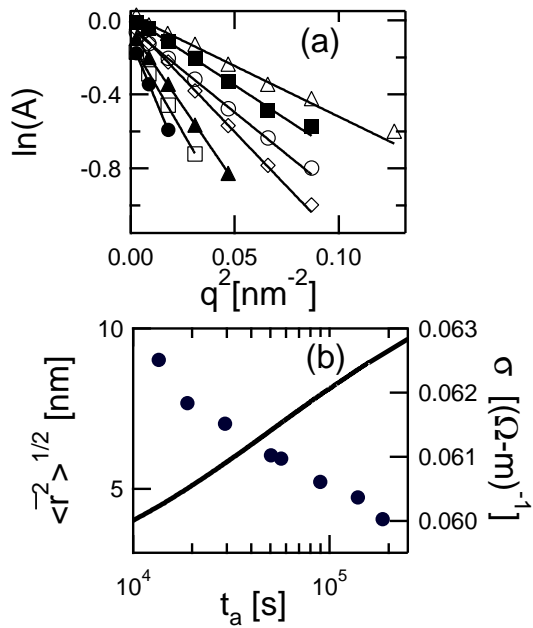}
 \caption{(a) The natural log of the
amplitude of $f(q,t)$ at short times ($t <$ 1 s) as a function of
$q^2$ at ages $t_a=1.3\times10^{4}$ s (solid circles),
$1.9\times10^{4}$ s (open squares), $3\times10^{4}$ s (solid
triangles), $5\times10^{4}$ s (open diamonds), $9\times10^{4}$ s
(open circles), $1.4\times10^{5}$ s (solid squares), and
$2\times10^{5}$ s (open triangles).  The solid lines show results
of fits to Eq.~(2).  (b) The root mean square displacement
$\langle\bar{r}^{2}\rangle^{1/2}$ (circles) for caged particle
motion and d.~c.~conductivity $\sigma$ (line) for a laponite
suspension of volume fraction $\phi = 0.012$ as function of age.}
\label{fig4}
\end{figure}

An increasing repulsion not only explains the increasing restriction in
caged motion but also provides a mechanism for the slow dynamics.  As
the internal stress grows with increasing repulsion, it will surpass the
local yield stress for mechanically weak arrangements of particles,
leading to local restructuring and strain like that pictured by Cipelletti
{\it et al.}  This connection between fast and slow local dynamics in
laponite thus demonstrates how aging does not always require quenched
disorder but rather can occur via the build-up of stress in an otherwise
annealed system.  Finally, we note that while the interactions were
predominantly repulsive for the solution conditions we
employed~\cite{repulsive,morvan}, laponite can form attractive fractal aggregates under
many conditions, such as with added salt ~\cite{nikolai,pignon}.  The
amplitude of fast dynamics in such aggregates should vary as
$\langle\bar{r}^{2}\rangle \sim 1/\kappa_0$, where $\kappa_0$ is the spring constant
of the attractive bond~\cite{krall}.  Increasing these particle
attractions would decrease $\langle\bar{r}^{2}\rangle$ and build internal
stress, possibly leading to a scenario that mirrors our observations for
the repulsive system.  This correspondence between attractive and
repulsive suspensions thus illustrates the potentially broad relevance of this
mechanism for aging, which relies not specifically on disorder but
rather on evolving interparticle interactions.

We thank B. Chung, D. Ho, S. Narayanan, and A. Sandy for their
assistance.  Funding was provided by the NSF (DMR-0134377 and DMR-0071755).
Acknowledgement is also made to the donors of The Petroleum Research
Fund, administered by the ACS.  Use of the APS was supported by the DOE,
Office of Basic Energy Sciences, under Contract No.~W-31-109-Eng-38.




\begin{references}
\bibitem{sigma_error}
Converting from impedance to $\sigma$ required scaling by a
multiplicative constant that had roughly 5\% uncertainty, based on
the capacitor geometry, but did not introduce error in the
variation of $\sigma$ with $t_a$.
\end{references}
\end{document}